\documentclass[twocolumn]{autart}    
\usepackage{xcolor}
\usepackage{subcaption} 
\usepackage{multirow}
\usepackage{algpseudocode} 
\usepackage{algorithm}
\usepackage{booktabs}  
\usepackage{graphicx}          
\usepackage{epstopdf}
\usepackage{amssymb}   
\usepackage{amsmath}
\usepackage{mathtools}
\makeatother

\newcommand{\bs}{\boldsymbol}

\newcommand{\off}{\mathrm{off}}
\newcommand{\true}{\mathrm{true}}

\algrenewcommand\algorithmicrequire{\textbf{Input:}}
\algrenewcommand\algorithmicensure{\textbf{Output:}}

\begin{document}

\begin{frontmatter}

\title{Online outlier-resilient data-driven moving horizon estimation for linear discrete-time systems\thanksref{footnoteinfo}} 

\thanks[footnoteinfo]{This research is supported by the Ministry of Education, Singapore, under its Academic Research Fund Tier 1 (RG95/24 \& RG63/22). *Corresponding author: Xunyuan Yin.}

\author[CCEB]{Jingshi Yao}\ead{jingshi001@e.ntu.edu.sg},
\author[NEWRI]{Zenghong Huang}\ead{zenghong.huang@ntu.edu.sg},
\author[CCEB]{Xiaojie Li}\ead{xiaojie002@e.ntu.edu.sg},
\author[CCEB,NEWRI]{Xunyuan Yin}\ead{xunyuan.yin@ntu.edu.sg}

\address[CCEB]{School of Chemistry, Chemical Engineering and Biotechnology, Nanyang Technological University, 62 Nanyang Drive, 637459, Singapore}  
\address[NEWRI]{Nanyang Environment and Water Research Institute, Nanyang Technological University, 1 CleanTech Loop, 637141, Singapore}


\begin{abstract}                         
{This paper addresses data-driven state estimation for linear systems of which online output measurements may be corrupted by sparse outliers of arbitrary magnitude. Using only offline input-state-output data without explicitly identifying a system model, we propose an outlier-resilient data-driven moving horizon estimation scheme, termed OORD-MHE, based on Willems' fundamental lemma.
The scheme employs a truncated-quadratic loss on the output residuals, parameterized by a truncation threshold, to mitigate the influence of online outliers. 
We establish sufficient conditions under which the resulting estimation error remains uniformly bounded.
Simulation results demonstrate that the proposed OORD-MHE scheme generates accurate state estimates in the presence of sparse online outliers.}
\end{abstract}

\end{frontmatter}

\section{Introduction}
State estimation plays a crucial role in system monitoring, diagnosis, and control~\cite{kalman1960new,mayne2014model}. {For linear systems,} the Kalman filter is a classical and widely adopted estimation method, valued for its recursive structure, computational efficiency, and guaranteed stability~\cite{kalman1960new,anderson2005optimal,rao2001constrained}.
However, the standard Kalman filter framework cannot explicitly handle state or input constraints~\cite{rao2001constrained}.
Moving horizon estimation (MHE) has been developed as an optimization-based state estimation approach formulated over a receding horizon~\cite{rao2001constrained,rawlings2020model}.
By explicitly incorporating system dynamics and inequality constraints into an online optimization problem, MHE can provide accurate and constraint-compliant state estimates~\cite{rao2001constrained,rao2003constrained}.

Most existing MHE schemes require an accurate dynamic model of the underlying system~\cite{rawlings2020model,rao2003constrained,alessandri2008moving,zou2020mhe,schiller2023lyapunov,yin2021event,schneider2017partition}.
{However, establishing an accurate mechanistic model can be difficult or costly~\cite{yan2025selftuning}.
A conventional indirect data-driven approach first identifies an explicit state-space model from offline data and then formulates an MHE scheme using the identified model~\cite{yan2025selftuning}. This type of approach can leverage the existing system identification~\cite{van2012subspace} and estimation methods~\cite{rawlings2020model,rao2003constrained,alessandri2008moving,zou2020mhe,schiller2023lyapunov,yin2021event,schneider2017partition}; meanwhile, the performance of this type of approach depends on modeling choices, including the model structure and order, identification objective, and regularization strategy~\cite{van2012subspace}.
An alternative is a direct data-driven approach that computes state estimates from trajectories without first identifying an explicit parametric model~\cite{wolff2024robust}. 
Based on Willems' fundamental lemma and relevant work~\cite{willems2005note,markovsky2008data,coulson2019data}, 
{\color{blue}a data-driven MHE framework was proposed in~\cite{wolff2024robust}, which} represents admissible system trajectories using offline input-state-output data while retaining the ability of MHE to incorporate constraints.} 

State estimation performance can be substantially degraded by online outliers, which represent observations that deviate significantly from the underlying data patterns due to sensor malfunctions or human error~\cite{liu2004line}.
Such outliers are frequently encountered across different areas, such as chemical processes~\cite{liu2004line}, health monitoring~\cite{chen2022gross}, and environmental monitoring~\cite{garces2011outliers}. 
Outliers are typically sporadic yet arbitrarily large, and even a single outlier may cause significant estimation {bias~\cite{duran2024outlier,cao2025robust}}.
{In this work, we aim to develop a data-driven MHE approach that can provide robust state estimates in the presence of online outliers without requiring an explicit system model.}

{Existing state estimation approaches capable of handling online outliers can typically be divided into two categories. The {\color{blue}first category} follows a detect-and-discard approach \cite{gibbs2013new,alessandri2016moving}.
For example, in the Kalman filtering context, statistical tests can be used to skip the measurement update when a test statistic exceeds a prescribed threshold~\cite{gibbs2013new}.
Similar ideas have been applied to MHE, where suspicious observations are discarded within each horizon while ensuring stability~\cite{alessandri2016moving}. 
The second category reduces the influence of online outliers through robust inference or robust penalties~\cite{akkaya2008robust,roth2013student,boncelet1983approach,gandhi2009robust,liu2017maximum,chen2017maximum}.}
When the contaminant distribution is known, an outlier-robust loss function can be formulated by replacing the Gaussian probability density function (PDF) in generalized Bayesian inference with the true contaminant PDF, such as the Student’s t-distribution~\cite{roth2013student} or a generalized logistic likelihood~\cite{akkaya2008robust}. In practice, however, the contaminant distribution is usually unknown, which motivates the development of nonparametric robust loss functions that can adapt to various types of outliers.
Commonly adopted nonparametric loss functions include the Huber loss, which retains a quadratic form near the origin but transitions to linear growth for large residuals~\cite{boncelet1983approach,gandhi2009robust}, and the correntropy loss, which is derived from the maximum correntropy criterion~\cite{liu2017maximum,chen2017maximum}.
When outlier magnitudes may be unbounded, it is appealing to employ a loss that caps the contribution of sufficiently large residuals. This observation motivates the truncated-quadratic penalty adopted in this paper.

In this paper, {we consider linear discrete-time systems of which online measurements may be corrupted by sparse outliers of arbitrary magnitude.}
Based on the data-driven MHE framework in~\cite{wolff2024robust}, we propose an online outlier-resilient data-driven moving horizon estimation (OORD-MHE) scheme, in which the output residuals in the cost function are penalized by a truncated-quadratic loss with a tunable truncation threshold. For linear systems, we establish uniform boundedness of the estimation error with suitable parameters and a sufficiently accurate initial prior, without requiring any upper bound on the outlier magnitude. To the best of our knowledge, such boundedness guarantees under arbitrary-magnitude outliers have not been previously reported in the state estimation literature, { even for model-based estimation methods}.
The main contributions of this work are summarized as follows:
\begin{enumerate}{
  \item We propose a data-driven OORD-MHE scheme for handling online measurement outliers without requiring explicit system identification.
  
  \item We establish uniform boundedness of the estimation error for linear systems subject to online outliers of arbitrary magnitude.
  
  \item The proposed method is evaluated on a linearized four-water-tank process; the simulation results demonstrate its robustness to online outliers.}
\end{enumerate}

The remainder of the paper is organized as follows. Section~\ref{section2} introduces the preliminaries and formulates the state estimation problem {in the presence of online outliers.
Section~\ref{section3} presents the proposed OORD-MHE and the associated boundedness analysis of the resulting estimation error.
Section~\ref{section5} reports simulation results.}
Finally, Section~\ref{section6} concludes the paper.

\section{Preliminaries and problem formulation}
\label{section2}
\subsection{{Notation}}
{Throughout the paper,}
$\mathbb{Z}$ denotes the set of all integers;
$\mathbb{Z}_{\geq a}$ denotes the set of integers greater than or equal to $a$; 
$\mathbb{Z}_{[a,b]}$ denotes the set of integers from $a$ to $b$ (inclusive). 
{ $L\ominus k :=\min\{L,k\}$.} 
Given a matrix $M\in \mathbb{R}^{m\times n}$, 
$\|M\|_{F} = \sqrt{\sum_{i}\sum_j M_{ij}^2}$ denotes the Frobenius norm.
For a symmetric matrix $M$, $M\succ0$ indicates that $M$ is positive definite;
{$\bar\lambda(M)$ and $\underline \lambda(M)$ denote its largest and smallest eigenvalues, respectively.}
For a vector $x\in\mathbb{R}^{n}$, $\|x\|=\sqrt{x^\top x}$ denotes its $\ell_2$-norm, and $\|x\|_M=\sqrt{x^\top Mx}$ denotes {\color{blue}the} weighted $\ell_2$-norm. 
For any finite set $\Gamma$, $|\Gamma|$ denotes its cardinality.
Given a finite sequence $\{x_i\}_{i=0}^{N-1} \coloneqq \{x_0, x_1, \cdots, x_{N-1}\}$, define its stacked column vector by $x_{[0,N-1]}=[x_0^\top,\ldots,x_{N-1}^\top]^\top$.
The Hankel matrix of $x_{[0,N-1]}$ with depth $L\in\mathbb{Z}_{[1,N]}$ is defined by
\[
    H_L(x_{[0,N-1]}) = 
    \begin{bmatrix}
        x_0 & x_1 & \cdots & x_{N - L} \\
        x_1 & x_2 & \cdots & x_{N - L + 1} \\
        \vdots & \vdots & \ddots & \vdots \\
        x_{L - 1} & x_{L} & \cdots & x_{N - 1}
    \end{bmatrix} {.}
\]

\subsection{Linear system and its behavioral representations}

Consider the following linear discrete-time system:
\begin{equation}
    \label{outliereq-sys}
    \begin{aligned}
        x_{k+1} &= A x_k + B u_k \\
        y_k &= C x_k + D u_k
    \end{aligned}
\end{equation}
where $x_k\in \mathbb{R}^{n_x}$, $u_k\in \mathbb{R}^{n_u}$, and $y_k\in \mathbb{R}^{n_y}$ denote the system state, known input, and system output at time instant $k\in\mathbb{Z}_{\geq 0}$, respectively; $A$, $B$, $C$, and $D$ are constant matrices of compatible dimensions.

The system in \eqref{outliereq-sys} can be represented in terms of its trajectories, without explicitly using the system matrices (i.e., $A$, $B$, $C$, and $D$)~\cite{willems2005note,berberich2020trajectory}; this representation relies on the concept of persistency of excitation, defined below.

\begin{defn}[\cite{willems2005note}]
    An input sequence $\{u_k\}_{k=0}^{N-1}$ is said to be persistently exciting of order $L$ if the rank of $H_L(u_{[0,N-1]})$ is $n_u L$.
\end{defn}

\begin{lem}[\cite{willems2005note}]
    \label{lemma-hankel}
    Consider an input-state trajectory $\{u_k\}_{k=0}^{N-1},\{x_k\}_{k=0}^{N-1}$ generated by the controllable system~\eqref{outliereq-sys}. If $\{u_k\}_{k=0}^{N-1}$ is persistently exciting of order $L+n_x$, then
            $\begin{bmatrix} 
                H_L(u_{[0,N-1]}) \\ 
                H_{1}(x_{[0,N-L]}) 
            \end{bmatrix}$ has full row rank. 
\end{lem}

\begin{lem}[\cite{willems2005note,berberich2020trajectory}]
    \label{lemma-willems}
    Consider an input-output trajectory $\{u_k\}_{k=0}^{N-1},~\{y_k\}_{k=0}^{N-1}$ generated by the controllable system~\eqref{outliereq-sys}. Assume $\{u_k\}_{k=0}^{N-1}$ is persistently exciting of order $L+n_x$. Then, $\{u_k'\}_{k=0}^{L-1},~\{y_k'\}_{k=0}^{L-1}$ constitute a trajectory of \eqref{outliereq-sys} if and only if there exists $g\in\mathbb{R}^{N-L+1}$ such that
            \begin{equation}
                \label{outliereq-thm}
                \begin{bmatrix} 
                    H_L(u_{[0,N-1]}) \\ 
                    H_L(y_{[0,N-1]})                     
                \end{bmatrix} g = 
                \begin{bmatrix} 
                    u_{[0,L-1]}' \\ 
                    y_{[0,L-1]}'                     
                \end{bmatrix} {.}
            \end{equation}
\end{lem}
Let $\{x_k\}_{k=0}^{N-1}$ and $\{{x}_k'\}_{k=0}^{L-1}$ represent the state trajectories of system~\eqref{outliereq-sys} corresponding to $\{u_k\}_{k=0}^{N-1},~\{y_k\}_{k=0}^{N-1}$ and $\{u_k'\}_{k=0}^{L-1},~\{y_k'\}_{k=0}^{L-1}$, respectively. Then, the following relationship holds \cite{berberich2020trajectory}:
\begin{equation}
    \label{outliereq-willems-x}
    H_{L}(x_{[0,N-1]})g = {x}_{[0,L-1]}'{.}
\end{equation}
According to Lemma~\ref{lemma-willems} and \eqref{outliereq-willems-x}, any admissible input-output (and corresponding state) trajectory of a controllable linear system of the form \eqref{outliereq-sys} can be expressed as a linear combination of columns of the offline Hankel matrix, provided that
the input trajectory is persistently exciting of a sufficiently high order.

\subsection{Problem formulation}
\label{sec-problem-formulation}

We adopt the data-driven state estimation framework in~\cite{wolff2024robust}, which comprises an offline phase and an online phase, and extend it to the case in which online measurements may be contaminated by outliers.

\textbf{Offline phase:}   
{Offline data are collected during this phase.
At} each sampling instant $k$, the offline input $u^{{\color{blue}\off}}_k$ of system~\eqref{outliereq-sys} is directly available, while the offline state and output measurements, denoted by $\tilde{x}^{{\color{blue}\off}}_k$ and $\tilde{y}^{{\color{blue}\off}}_k$, are corrupted by bounded measurement noise. Accordingly, these corrupted offline state and output measurements are modeled as{:
\begin{equation}
    \label{outliereq-contaminant}
    \begin{aligned}
        \tilde{x}^{{\color{blue}\off}}_k &= x^{{\color{blue}\off}}_k + \epsilon^{{\color{blue}\off}}_{x,k}  \\
        \tilde{y}^{{\color{blue}\off}}_k &= y^{{\color{blue}\off}}_k + \epsilon^{{\color{blue}\off}}_{y,k} 
    \end{aligned}
\end{equation}
w}here $x^{{\color{blue}\off}}_k$ and $y^{{\color{blue}\off}}_k$ denote the bounded, noise-free offline state and output of system~\eqref{outliereq-sys}, respectively; 
$\epsilon^{{\color{blue}\off}}_{x,k}$ and $\epsilon^{{\color{blue}\off}}_{y,k}$ denote the bounded offline state and output measurement noise satisfying $\| \epsilon^{{\color{blue}\off}}_{x,k} \| \leq  \bar \epsilon^{{\color{blue}\off}}_{x}, ~\| \epsilon^{{\color{blue}\off}}_{y,k} \| \leq  \bar \epsilon^{{\color{blue}\off}}_y$ with $\bar \epsilon^{{\color{blue}\off}}_{x}, ~\bar \epsilon^{{\color{blue}\off}}_y > 0$.
Consequently, the available offline dataset of length $N^{{\color{blue}\off}}$ contains
$u^{{\color{blue}\off}}_{[0,N^{{\color{blue}\off}}-1]}$, $\tilde{y}^{{\color{blue}\off}}_{[0,N^{{\color{blue}\off}}-1]}$, and $\tilde{x}^{{\color{blue}\off}}_{[0,N^{{\color{blue}\off}}]}$.
For convenience, we define the overall noise bound as $\bar{\epsilon}^{{\color{blue}\off}} \coloneqq \max\{\bar \epsilon^{{\color{blue}\off}}_{x},\bar \epsilon^{{\color{blue}\off}}_y\}$.

While full-state information may be unavailable in real time and needs to be estimated during the online phase, we assume that the state measurements can be obtained offline, possibly with time delays and at additional cost (e.g., due to the use of high-grade sensors or laboratory analyzers).
For example, in chemical processes, substance concentrations are often measured offline using specialized analytical instruments, and the results may become available after a significant delay~\cite{nicoletti2009computational,bourgeois2001line}.
This setting, where state measurements are available only as historical data, aligns with the case considered in~\cite{wolff2024robust}.

\textbf{Online phase:}  
{State estimation is performed during this phase.
At} each sampling instant $k$, the online input $u_k$ of system~\eqref{outliereq-sys} is directly available, whereas the online output measurement $\tilde{y}_k$ {is affected by two types of corruption: (i) noise, which is prevalent and bounded; (ii) outliers, which are sparse and may have arbitrarily large magnitudes. Accordingly, the corrupted online output measurement is modeled as}
\begin{equation}
    \label{outliereq-tilde-y}
    \tilde{y}_k = y_k + \epsilon_k + o_k
\end{equation}
where $y_k$ denotes the noise-free online output of system \eqref{outliereq-sys}; $\epsilon_k$ denotes the bounded online output noise satisfying $\|\epsilon_k\| \leq  \bar \epsilon$ with $\bar \epsilon>0$; and $o_k$ denotes the sparse online outlier. The system state $x_k$ is unmeasurable during the online phase.

\begin{prob}
    \label{problem}
    For system~\eqref{outliereq-sys}, given (i) offline data $u^{{\color{blue}\off}}_{[0,N^{{\color{blue}\off}}-1]}$, $\tilde{y}^{{\color{blue}\off}}_{[0,N^{{\color{blue}\off}}-1]}$, and $\tilde{x}^{{\color{blue}\off}}_{[0,N^{{\color{blue}\off}}]}$;
    (ii) an initial prior estimate $\bar{x}_0$;
    (iii) the online input trajectory and measured output trajectory,
    the objective is to estimate the system state at each sampling instant during the online phase, {despite outlier contamination of the online output measurements.}
\end{prob}

\begin{rem}
The offline measurements can be collected using laboratory analyzers or high-fidelity sensors~\cite{nicoletti2009computational,bourgeois2001line}, and they are generally of higher quality than online measurements. Moreover, the offline data can often be preprocessed prior to online state estimation. 
{
When offline outliers are suspected, a preprocessing method such as robust principal component analysis~\cite{candes2011robust} may be applied. A rigorous
analysis of how such preprocessing affects the error bound of the subsequent OORD-MHE is beyond the scope of the current work.}
\end{rem}

\section{Online outlier-resilient data-driven MHE}
\label{section3}

\subsection{Formulation of the online outlier-resilient data-driven MHE}
\label{section-OORD-MHE}

{
We develop the state estimation scheme based on the data-driven MHE framework in~\cite{wolff2024robust}, 
in which the system trajectories are parameterized by Hankel matrices constructed from
offline input--output--state data, and the output residuals are penalized by a
quadratic loss. 
Such a quadratic penalty is sensitive to large online
measurement outliers, since a single large residual may dominate the objective
function and subsequently bias the resulting state estimate. To reduce the influence of
such outliers, we replace the quadratic output-residual penalty with the
following truncated-quadratic loss~\cite{huber2011robust}:
}
\begin{equation}
    \label{outliereq-loss-sigma-y}
    \ell(\sigma^y,\bar l)=\min\{\|\sigma^y\|^2,\bar l\},
\end{equation}
{
where $\bar l>0$ is the truncation threshold. The loss in \eqref{outliereq-loss-sigma-y} is quadratic near
the origin and saturates for large residuals. Consequently, small nominal residuals
are penalized in a similar manner as in the standard quadratic MHE cost, whereas large residuals caused by online outliers have a bounded contribution to the objective. Robust loss functions of this type are commonly used to limit the influence of outliers in estimation problems~\cite{duran2024outlier,huber2011robust}.

Following~\cite{wolff2024robust}, we formulate the online outlier-resilient data-driven MHE (OORD-MHE) based on the non-parametric representation theory \cite{willems2005note,berberich2020trajectory}, which requires the system to be controllable.

\begin{assum}
    \label{assumption-controllable}
    System~\eqref{outliereq-sys} is controllable.
\end{assum}}

At each time instant $k\in\mathbb Z_{\geq 1}$, the OORD-MHE uses the available online data $u_{[{l_k},k-1]}$, $\tilde y_{[{l_k},k-1]}$, and the prior estimate
$\bar x_{{l_k}}$. The OORD-MHE then solves the following optimization problem:
\begin{subequations}
\label{outliereq-OORD-MHE}
\begin{align}
    & \min_{\substack{g_k,\,{\hat{\bs{x}}_k,\,
    \bs{\sigma}^x_k,\,
    \bs{\sigma}^y_k}}}
    \quad  J_k \label{outliereq-OORD-MHE-obj} \\
    \mathrm{s.t.}
    &\begin{bmatrix}
        H_{{L\ominus k}}(u^{{\color{blue}\off}}_{[0,N^{{\color{blue}\off}}-1]}) \\
        H_{{L\ominus k}}(\tilde y^{{\color{blue}\off}}_{[0,N^{{\color{blue}\off}}-1]}) \\
        H_{{L\ominus k}+1}(\tilde x^{{\color{blue}\off}}_{[0,N^{{\color{blue}\off}}]})
    \end{bmatrix}g_k
    \!=\!
    \begin{bmatrix}
        u_{[{l_k},k-1]} \\
        \tilde y_{[{l_k},k-1]}-{\bs{\sigma}^y_k} \\
        {\hat{\bs{x}}_k+\bs{\sigma}^x_k}
    \end{bmatrix}\!
    \label{outliereq-OORD-MHE-hankel} \\
    &~\hat x_{j|k}\in\mathbb X,\quad \forall j\in\mathbb Z_{[{l_k},k]}
    \label{outliereq-constraint} \\
    &~\|g_k\|\leq \bar g
    \label{outliereq-g-max}
\end{align}
\end{subequations}
where the cost function $J_k$ is defined as
\begin{align}
        J_k 
        = &\rho^{{L\ominus k}} \| \hat{x}_{{l_k}|k} - \bar{x}_{{l_k}} \|^2_{P}
        + {\color{blue}\mu_x} \| {\bs{\sigma}^x_k} \|^2  \notag\\
        & + {\color{blue}\mu_g}\bar{H}_\epsilon^2\| g_k \|^2  
        + \sum_{j={l_k}}^{k-1}\ell({\sigma}^y_{j|k}, \bar{l}) {.}
    \label{outliereq-loss-OORD-MHE}
\end{align}
{
The estimation horizon is $L\in\mathbb{Z}_{\geq 1}$.
The index $l_k:=k-(L\ominus k)$ denotes the left endpoint of the estimation horizon at time instant $k$.} In \eqref{outliereq-OORD-MHE},
the decision variable {$\hat{\bs{x}}_k := \hat x_{[{l_k},k]|k}$} denotes the sequence of state estimates from time instant ${l_k}$ to $k$, calculated at time instant $k$;
{$\bs{\sigma}^y_k:=\sigma^y_{[{l_k},k-1]|k}$} is the sequence of output residuals from time instant ${l_k}$ to {\color{blue}$k-1$}, calculated at time instant $k$, capturing offline and online output corruption;
{$\bs{\sigma}^x_k:=\sigma^x_{[{l_k},k]|k}$} consists of slack variables from time instant ${l_k}$ to $k$, calculated at time instant $k$, introduced to account for offline state noise.
Set $\mathbb{X}$ specifies the state constraints.
$\bar g>0$ is a scalar tuning parameter for solution stability, as will be detailed in Section~\ref{sec:stability}.
In the cost function~\eqref{outliereq-loss-OORD-MHE}, 
$P\succ 0$, $0<\rho<1$, ${\color{blue}\mu_x},{\color{blue}\mu_g}>0$, $\bar{H}_\epsilon = \sqrt{N^{{\color{blue}\off}}}\bar{\epsilon}^{\color{blue}\off}$, and $\bar l>0$ are constant parameters.
The global optimizer of OORD-MHE in \eqref{outliereq-OORD-MHE} is denoted by $(g_k^*,\hat{\bs{x}}_k^*,\bs{\sigma}^{x, *}_k,\bs{\sigma}^{y, *}_k)$. 
{\color{blue} The prior estimate for \eqref{outliereq-OORD-MHE} at time instant $k\in\mathbb{Z}_{\geq1}$ is $\bar{x}_{{l_k}}=\hat{x}^*_{{l_k}|{l_k}}$, with the initialization $\hat{x}^*_{0|0}:=\bar{x}_0$ and inital state prior ${\bar x}_0$.}

\begin{rem}
\label{remark:solution}
The variables $g_k$, $\hat{\bs{x}}_k$, $\bs{\sigma}^{x}_k$, and $\bs{\sigma}^y_k$ are explicitly listed in the OORD-MHE design in \eqref{outliereq-OORD-MHE} as decision variables. 
Meanwhile, we note that once $g_k$ and $\hat{\bs{x}}_k$ are specified, the corresponding state slack variable $\bs{\sigma}^{x}_k$ and output slack variable $\bs{\sigma}^y_k$ are uniquely determined by the
equality constraint \eqref{outliereq-OORD-MHE-hankel}. 
\end{rem}

\begin{rem}
The proposed OORD-MHE in \eqref{outliereq-OORD-MHE} is developed based on the direct data-driven framework in~\cite{wolff2024robust}. Within the context of predictive control, it has been shown that, for the formulations being considered, certain regularized direct methods can be viewed as convex relaxations of the corresponding indirect methods, with regularization accounting for an implicit identification step \cite{dorflerref}.
We note that the adoption of the direct method in this work is not because of a claim that the direct approach universally outperforms the indirect alternatives. Instead, following \cite{wolff2024robust}, it avoids separate parametric model identification. 
\end{rem}

\begin{rem}
Stability analyses developed for conventional quadratic objectives cannot be directly extended to estimators utilizing robust loss functions, including the truncated-quadratic loss in \eqref{outliereq-loss-sigma-y}. Existing stability results for estimators with robust losses often require an explicit bound on the outlier magnitude~\cite{alessandri2016moving}, or only establish that the estimation error does not deteriorate abruptly following the occurrence of outliers~\cite{duran2024outlier}.
In contrast, the subsequent analysis shows that, under suitable initial conditions, the estimation error of the proposed data-driven MHE remains uniformly bounded over time without requiring a prescribed upper bound on the magnitude of the online outliers.
\end{rem}

\subsection{Stability analysis for OORD-MHE}
\label{sec:stability}
In this subsection, we prove that the estimation error of the proposed OORD-MHE scheme in \eqref{outliereq-OORD-MHE} is uniformly bounded,
with the bound growing at most exponentially in the maximum number of outliers allowed within each estimation horizon. 

For system~\eqref{outliereq-sys}, let $x(k|x_0, u_{[0,k]})$ and $y(k|x_0, u_{[0,k]})$ denote the state and output at time instant $k$ corresponding to the initial state $x_0$ and input sequence $\{u_i\}_{i=0}^k$, respectively.
We recall the definition of incremental exponential uniform output-to-state stability (e-UOSS) introduced in~\cite{wolff2024robust}.
\begin{defn}[{\cite[Definition~2]{wolff2024robust}}]
    \label{def-e-uoss}
    The system~\eqref{outliereq-sys} is said to be e-UOSS if there exist constants $p_0 \geq 1,~p_1\geq 0$, and $0<\eta<1$, such that for every pair of initial states $x_0,~x_0'\in \mathbb{R}^{n_x}$ and any input sequence $\{u_i\}_{i=0}^k$, the following inequality holds for all $k\in \mathbb{Z}_{\geq 0}$:
    \begin{align}
        \label{outliereq-e-uoss}
        &\| x(k|x_0, u_{[0,k]}) - x(k|x_0', u_{[0,k]}) \| 
            \leq p_0 \eta^k \| x_0 - x_0' \|   \notag \\
            &\quad+\sum_{j=0}^{k-1} p_1 \eta^{k - j - 1} \| y(j|x_0, u_{[0,j]}) -  y(j|x_0', u_{[0,j]}) \|{.}
    \end{align}
\end{defn}

Definition~\ref{def-e-uoss} establishes a bound on the state deviation arising from discrepancies in the initial states and output trajectories. {In the following, we introduce the additional assumptions and notations needed for the stability analysis.}

\begin{assum}
    \label{assumption-detectable}
    System~\eqref{outliereq-sys} has the e-UOSS property.
\end{assum}
{\color{blue}We noted that for LTI system \eqref{outliereq-sys}, the e-UOSS property is equivalent to detectability \cite{knufer2020time}.}

\begin{assum}
    \label{assumption-PE}
    The offline input trajectory $\{u^{{\color{blue}\off}}_k\}_{k=0}^{N^{{\color{blue}\off}}-1}$ of system~\eqref{outliereq-sys} is persistently exciting (PE) of order $L + n_x + 1$.
\end{assum}

\begin{assum}
    \label{assumption-trajectory-bound}
    The online input-state trajectory of system~\eqref{outliereq-sys} evolves within compact sets; that is, there exist constants $\bar{x}, \bar{u} \geq 0$ such that the true online trajectory of system~\eqref{outliereq-sys} satisfies 
    \[
    x_k\in\mathbb X,\quad
    \|x_k\|\leq\bar x,\quad
    \|u_k\|\leq\bar u,
    \quad
    \forall k\in\mathbb Z_{\geq0}.
    \]
\end{assum}

Since the truncated-quadratic loss \eqref{outliereq-loss-sigma-y} is employed for the output residuals, the stability of the OORD-MHE in \eqref{outliereq-OORD-MHE} depends primarily on the number of outliers rather than their magnitude, even if individual residuals may be arbitrarily large. 
Let $\Gamma_k^o$ denote the set of time instants within the past estimation horizon of time instant $k$, at which the combined effect of online measurement noise and outliers exceeds the nominal noise bound:
\begin{equation}
    \label{outliereq-define-gamma-1}
    \Gamma_k^o \coloneq \{j\in \mathbb{Z}_{[{l_k},k-1]} \mid \| \epsilon_j+o_j \|^2 > \bar{\epsilon}^2\}{.}
\end{equation}

{
\begin{assum}
    \label{assumption-outlier-number}
    There exists an integer $n_o\in\mathbb{Z}_{\geq0}$, such that 
    \[
    |\Gamma_k^o|\leq n_o,
    \qquad
    \forall k\in\mathbb Z_{\geq1}.
    \]
\end{assum}
}

To facilitate stability analysis, we further define $\Gamma^*_{k, \delta}$ as the set of time instants within the past estimation horizon of time instant $k$ at which {\color{blue}the squared norm of the optimal output residual is at least} a prescribed threshold $\delta\geq 0$:
\begin{equation}
    \label{outliereq-define-gamma-2}
    \Gamma^*_{k, \delta} \coloneq \{j\in \mathbb{Z}_{[{l_k},k-1]} \mid \| \sigma^{y,*}_{j|k} \|^2\geq \delta \}{.}
\end{equation}

\begin{prop}
    \label{prop-number}
    {
    Consider system~\eqref{outliereq-sys} and the OORD-MHE in~\eqref{outliereq-OORD-MHE}. Suppose that
    Assumptions~\ref{assumption-controllable}--\ref{assumption-trajectory-bound} hold and $\bar\epsilon>0$. Given $L\in\mathbb{Z}_{\geq 1}$, define
    \begin{equation}
        \label{alpha1}
        \begin{aligned}
        \alpha_1(L,z,s)
        :={}&
        (2L+1)\bar\epsilon^2
        +\bar\lambda(P)\rho^Ls^2\\
        &+
        \bigl({\color{blue}\mu_x}(L+1)+{\color{blue}\mu_g}+2L\bigr)z^2,
        \end{aligned}
    \end{equation}
    and
    \begin{equation}
    \label{outliereq-define-gmax}
    {G(L)} := \max_{j\in\mathbb Z_{[1, L]}}
    \left\| 
        \begin{bmatrix} 
            H_j(u^{{\color{blue}\off}}_{[0,N^{{\color{blue}\off}}-1]}) \\ 
            H_1(x^{{\color{blue}\off}}_{[0,N^{{\color{blue}\off}}-j]}) 
        \end{bmatrix}^\dagger
    \right\| \sqrt{j\bar{u}^2+\bar{x}^2}.
    \end{equation}
    If the following conditions hold: i) $\bar g\geq G(L)$, ii) $\bar s\geq0$, iii) $\bar l$ satisfies
    \begin{equation}
    \label{outliereq-bar-l}
    a\leq\bar l\leq a+\bar\epsilon^2,
    \end{equation}
    where}
    \begin{equation}
    \label{outliereq-a-definition}
    {a:=
    \alpha_1(L,\bar H_\epsilon\bar g,\bar s)},
    \end{equation}
    and iv) the prior estimation error satisfies
    \begin{equation}
    \label{outliereq-prior-error}
    \|\bar x_{l_k}-x_{l_k}\|
    \leq
    \sqrt{\rho}^{L-{L\ominus k}}\bar s,
    \end{equation}
    then the following relation is satisfied:
    \begin{equation}
    \label{outliereq-set-relation}
    |\Gamma_{k,a}^*|
    \leq
    2|\Gamma_k^o|.
    \end{equation}
    \end{prop}
Proposition~\ref{prop-number} establishes a relationship between $\Gamma_k^o$ and $\Gamma^*_{k, \delta}$,
thereby linking the occurrence of outliers to the magnitude of the optimal output residuals. 
The proof is provided in Appendix \ref{appendix2}.

For convenience, we define the following sets:
\begin{subequations}    
    \label{outliereq-gamma-derive}
    \begin{align}
        \Gamma_{k, a} & \coloneqq \Gamma_k^o \cup \Gamma^*_{k, a}{,}\\
        \bar{\Gamma}_{k, a} & \coloneqq \mathbb{Z}_{[{l_k},k-1]}\setminus {\Gamma}_{k, a}{.}
    \end{align}
\end{subequations}
Then, \eqref{outliereq-set-relation} and \eqref{outliereq-gamma-derive} lead to
\begin{equation}
    \label{outliereq-set-relation2}
    |\Gamma_{k, a}| \leq 3 |\Gamma_k^o  |{.}
\end{equation}
In the following, \ we establish an error-bound propagation inequality over one estimation horizon.  
\begin{thm}
    \label{theorem-one-horizon}
    Consider system~\eqref{outliereq-sys} and the OORD-MHE in~\eqref{outliereq-OORD-MHE}. Suppose that Assumptions~\ref{assumption-controllable}--\ref{assumption-outlier-number} hold, {$0<\eta<\rho<1$, and $\bar\epsilon>0$.
    Given $L\in\mathbb{Z}_{\geq 1}$, 
    for any $\bar g\geq G(L), ~\bar s\geq 0$ and $\bar l$ satisfying \eqref{outliereq-bar-l} with $a=\alpha_1(L,\bar H_\epsilon\bar g,\bar s)$,} if the prior error $\|\bar x_{l_k}-x_{l_k}\|$ satisfies \eqref{outliereq-prior-error}, then the following inequality holds
    \begin{align}
        \label{outliereq-one-step}
        &\| x_k-\bar{x}_k\| \leq \gamma_1(L, |\Gamma_{k, a}|) \sqrt{\rho}^L\bar{s}
            +\gamma_2(L, |\Gamma_{k, a}|, \bar{H}_\epsilon{\bar g}) \notag \\
        &  \qquad  +  \gamma_3(L, |\Gamma_{k, a}|)\sum_{j\in \bar{\Gamma}_{k, a}}  \| \epsilon_j+o_j \|
    \end{align}
    where $\gamma_1,\gamma_2,\gamma_3$ are non-negative functions {that are non-decreasing with respect to their second argument, grow at most polynomially with respect to their remaining arguments,} and satisfy $\gamma_2(L,0,0)=0$.
\end{thm}
The proof of Theorem~\ref{theorem-one-horizon} is provided in~Appendix \ref{appendix3}.
Theorem~\ref{theorem-one-horizon} demonstrates that the estimation error bound at time instant $k$ can be decomposed into three components: 
(i)~{ a prior-error-related term}, corresponding to  the first term on the right-hand side of \eqref{outliereq-one-step};
(ii)~a term reflecting the influence of the offline noise and online outliers, corresponding to the second term on the right-hand side of \eqref{outliereq-one-step};
(iii)~the accumulated effect of noise and outliers within time instant set $\bar{\Gamma}_{k, a}$, corresponding to  the third term on the right-hand side of \eqref{outliereq-one-step}.
Building on Theorem~\ref{theorem-one-horizon}, {Corollary~\ref{corollary-uniform-bound}} further establishes a uniform bound on the estimation error.

\begin{cor}
    \label{corollary-uniform-bound}
    Consider system~\eqref{outliereq-sys} and the OORD-MHE in~\eqref{outliereq-OORD-MHE}. 
    {
    Suppose Assumptions~\ref{assumption-controllable}--\ref{assumption-outlier-number} hold, $\bar\epsilon>0$, and $0<\eta<\rho<1$. Given $L\in\mathbb{Z}_{\geq 1}$, define
    \[
    \begin{aligned}
        & s(L, n_o, \bar{H}_\epsilon\bar g) := \dfrac{\gamma_2(L, 3n_o, \bar{H}_\epsilon\bar g)+\gamma_3(L, 3n_o) L \bar{\epsilon}}{1 - \gamma_1(L, 3n_o)\sqrt{\rho}^L}.
    \end{aligned}
    \]
    If $\bar g\geq G(L), ~\bar s\geq s(L, n_o, \bar{H}_\epsilon\bar g),~\|\bar{x}_{0} - x_{0} \| \leq \sqrt{\rho}^{L}\bar{s}$, $\bar l$ {\color{blue}satisfies} \eqref{outliereq-bar-l} with $a=\alpha_1(L,\bar H_\epsilon\bar g,\bar s)$, and
    \[
    \gamma_1(L, 3n_o)\sqrt{\rho}^L < 1,~ 
    \]} then the estimation error is uniformly bounded by $\bar s$:
    \begin{equation}
    \label{outliereq-uniform-bound}
        \|x_k-\bar x_k\|\;\le\;\bar{s}, \quad \forall k\in\mathbb{Z}_{\geq 0}{.}
    \end{equation}
\end{cor}

The proof of {Corollary~\ref{corollary-uniform-bound}} is provided in Appendix~\ref{appendix4}. 
{Corollary~\ref{corollary-uniform-bound}} shows that the state estimation error is uniformly bounded over time, provided that the parameters of OORD-MHE are properly chosen and the prior estimation error is sufficiently small.

\begin{rem}
    {
    In Corollary \ref{corollary-uniform-bound}, a sufficient contraction condition is $\gamma_1(L, 3n_o)\sqrt{\rho}^L < 1$, where
    $n_o$ is independent of $L$. 
    For fixed $n_o$, $\gamma_1(L,3n_o)$ grows at most polynomially in $L$, whereas $\sqrt{\rho}^L$ decays exponentially. Therefore, the condition holds for all sufficiently large admissible horizons $L$. Note that the PE requirement imposes the necessary data-length condition $N^{{\color{blue}\off}}\geq (L+n_x+1)(n_u+1)-1$. Consequently, increasing $L$ requires collecting a {\color{blue}correspondingly} longer PE offline trajectory.
    }
\end{rem}

\begin{rem}
    Corollary~\ref{corollary-uniform-bound} establishes uniform boundedness of the estimation error given by the proposed OORD-MHE. This result does not follow directly from Assumption~\ref{assumption-trajectory-bound} or the state constraint; Assumption~\ref{assumption-trajectory-bound} concerns the boundedness of the actual online trajectory, while the state constraint only leads to a potentially conservative bound induced by the constraint set $\mathbb X$. In contrast, the bound in Corollary~\ref{corollary-uniform-bound} explicitly depends on the estimator parameters, noise, and outliers. We do not claim exponential stability. Due to the truncated output-residual loss, large residuals are saturated rather than increasingly penalized, and the measurements may fail to provide a global correction mechanism when the prior estimate is far from the true state. Therefore, under the present assumptions, Corollary \ref{corollary-uniform-bound} is local in nature and does not imply global or semi-global exponential stability.
\end{rem}

\section{Case study}
\label{section5}
This section illustrates the effectiveness and outlier resilience of the proposed { OORD-MHE method in \eqref{outliereq-OORD-MHE}. }
Following~\cite{cao2025robust}, the outlier amplitude in the simulations is set to the maximum measurable output value of the corresponding sensors.
Here, we consider an open-loop stable linearized four-water-tank process (LFWTP), with the linearized system matrices given below~\cite{wolff2024robust}:
\begin{align}
    \label{outliereq-LFWTP}
    &A \!=\! 
        \begin{bmatrix}
            0.921 & 0     & 0.041 & 0 \\
            0     & 0.918 & 0     & 0.033\\
            0     & 0     & 0.924 & 0     \\
            0     & 0     & 0     & 0.937
        \end{bmatrix}\!,~
    B \!=\!
        \begin{bmatrix}
            0.017 & 0.001\\
            0.001 & 0.023\\
            0      & 0.061\\
            0.072 & 0
        \end{bmatrix}\!, \notag \\
    &C \!=\!
        \begin{bmatrix}
            1 & 0 & 0 & 0\\
            0 & 1 & 0 & 0
        \end{bmatrix}\!,~
    D \!=\! \begin{bmatrix}
             0 & 0\\
             0 & 0
        \end{bmatrix} {.}
\end{align}
The model in \eqref{outliereq-LFWTP} is used to generate data. 
An offline trajectory of length $N^{{\color{blue}\off}}=100$ is collected.
Each component of the offline input is independently sampled from the uniform distribution $\mathcal{U}(0,10)$. 
For each offline state and output component, the measurement noise is independently sampled from the normal distribution~$\mathcal{N}(0, 0.001^2)$ truncated to  $[-0.003, 0.003]$.

The data-driven MHE in \cite{wolff2024robust}, referred to as D-MHE~\cite{wolff2024robust}, is considered as the baseline. We set $L=7$, $\rho=0.95 ~(\rho>\eta)$, and $\bar g=100 ~(\bar g\geq G(L))$, where the values of $\eta$ and $G(L)$ are estimated from data. The arrival cost matrix for D-MHE~\cite{wolff2024robust} is denoted by $P^{ref}$ to distinguish it from that of OORD-MHE. 
{\color{blue}We choose three weighting parameter configurations $\mathcal C_1$, $\mathcal C_2$, and $\mathcal C_3$ for the LFWTP simulation as shown in Table \ref{table-configuration}, where $c_g=c_\alpha(\bar\epsilon_x^d+\bar\epsilon_y^d)$ and $c_\alpha$, $\bar\epsilon_x^d$, and $\bar\epsilon_y^d$ are defined in~\cite{wolff2024robust} for D-MHE.}
The first two D-MHE configurations are taken from \cite{wolff2024robust}. The weighting parameters for OORD-MHE are matched with D-MHE~\cite{wolff2024robust} since the output-residual weight of OORD-MHE is fixed to $1$. We empirically consider $\bar l\in\{1,100\}$.
The state constraint set is $\mathbb{X}=\{x\in\mathbb{R}^4|x\geq [0, 0, 0, 0]^\top\}$, and the initial state prior is set to $\bar{x}_0=[1, 2, 1, 2]^\top$. {All the above MHE problems are formulated in CasADi~\cite{andersson2019casadi} and solved using IPOPT~\cite{wachter2006implementation} with a zero initial guess. The maximum number of iterations is set to $1000$, and the termination tolerance is set to $10^{-10}$.

\begin{table}[t]{\color{blue}
\centering
\caption{Weighting parameter configurations for LFWTP}
\begin{tabular}{lccccccc}
\hline & $P^{ref}$ & $c_{\sigma^x}$ & $R$ & $c_g$ & $P$ & ${\color{blue}\mu_x}$ & ${\color{blue}\mu_g}\bar H_\epsilon^2$ \\
\hline
$\mathcal C_1$
    & $500 I_4$
    & $600$
    & $500 I_2$
    & $12$ 
    & $I_4$
    & $1.2$
    & $0.024$\\
$\mathcal C_2$ 
    & $500 I_4$
    & $600$
    & $100 I_2$
    & $12$
    & $5I_4$
    & $6$
    & $0.12$\\
$\mathcal C_3$ 
    & $500 I_4$
    & $600$
    & $500 I_2$
    & $120$
    & $I_4$
    & $1.2$
    & $0.24$\\
\hline
\end{tabular}
\label{table-configuration}}
\end{table}

\begin{table}[t]
\centering
\caption{{Mean squared errors (MSEs) of the state estimates for the LFWTP system under online outliers with probability $0.02$.}}
\begin{tabular}{lccc}
\hline
& $\mathcal C_1$
& $\mathcal C_2$
& $\mathcal C_3$\\
\hline
D-MHE~\cite{wolff2024robust}
& $0.7251$
& $0.7410$
& $0.7168$\\
OORD-MHE ($\bar l=1$)
& $1.1141$
& $1.2005$
& $1.1260$\\
OORD-MHE ($\bar l=100$)
& $\mathbf{0.5743}$
& $\mathbf{0.6986}$
& $\mathbf{0.5732}$\\
\hline
\end{tabular}
\label{table-compare}
\end{table}

\begin{figure}
    \centering
        \includegraphics[width=0.49\textwidth, page=1]{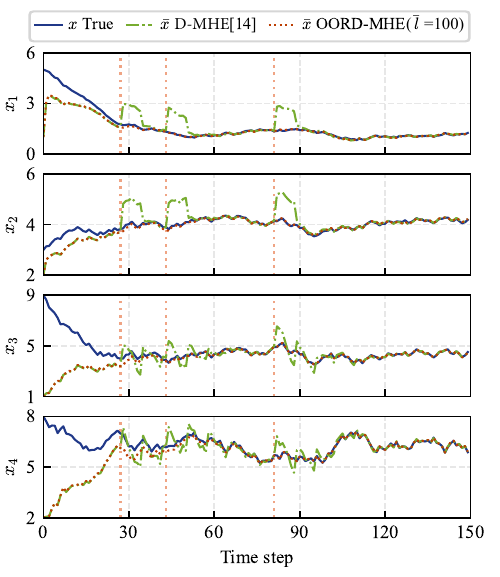}
    \caption{
    {State estimates from one Monte Carlo run under configuration $\mathcal{C}_1$ for OORD-MHE ($\bar l=100$) and D-MHE~\cite{wolff2024robust}.
    Vertical dashed lines indicate the time instants when outliers occur;
    solid trajectories correspond to the actual states; dash-dotted trajectories correspond to the estimates of D-MHE~\cite{wolff2024robust}; dotted trajectories correspond to the estimates of OORD-MHE ($\bar l=100$).  }    }
    \label{fig-sys1}
\end{figure}

During the online state estimation, 50 independent simulation runs of length $N=150$ are conducted.
Each component of the initial state is independently sampled from $\mathcal U(0,10)$. {\color{blue}The noise in each online output component} is independently sampled from $\mathcal N(0,0.5^2)$, and is then truncated to $[-1.5,1.5]$. Each online input component follows  $\mathcal{U}(0, 10)$.
At each time instant, the occurrence of online outliers is determined by an independent Bernoulli random variable with probability of 0.02. When an outlier occurs, a fixed value of 15 is added to the output measurements.
The estimation performance is evaluated using the mean squared error (MSE): $\frac{1}{50N}\sum_{i=1}^{50} \sum_{j=1}^{N} \|x^{(i)}_{j}-\bar x^{(i)}_{j}\|^2$.
Table~\ref{table-compare} reports the MSEs for the three parameter configurations. 
With $\bar l=1$, OORD-MHE has a larger MSE than D-MHE~\cite{wolff2024robust}. 
This behavior is consistent with the role of $\sqrt{\bar l}$ as the saturation threshold. Specifically, the chosen threshold $1$ is smaller than the maximum norm of the online noise, $1.5\sqrt{2}$, and the initial estimation error is not sufficiently small.

With $\bar l=100$, $\sqrt{\bar l}$ exceeds the noise level but is still below the outlier magnitude. Therefore, OORD-MHE
limits the influence of the outlier-contaminated residuals while avoiding frequent saturation of nominal residuals. For the three tested
configurations, OORD-MHE reduces the MSE compared to D-MHE~\cite{wolff2024robust} by $20.8\%$, $5.7\%$, and $20.0\%$, respectively.
 Fig.~\ref{fig-sys1} presents the state estimates from one Monte Carlo run under configuration $\mathcal C_1$ for OORD-MHE ($\bar l=100$) and D-MHE~\cite{wolff2024robust}. The state estimates of the D-MHE~\cite{wolff2024robust} exhibit pronounced deviations after the occurrences of outliers, whereas OORD-MHE gives smoother and more accurate estimates, which demonstrates its improved resilience to the outliers.}

\section{Conclusion}
\label{section6}
{We addressed the problem of data-driven moving horizon estimation for linear discrete-time systems whose online measurements may be corrupted by outliers. 
By introducing a truncated-quadratic loss for the output residuals, the proposed OORD-MHE approach mitigates the influence of sparse outliers of arbitrary magnitude.
We established a relationship between the outlier-contaminated time instants represented by $\Gamma_k^o$ and the large output residuals represented by $\Gamma_{k,a}^*$ for the optimal OORD-MHE solution. Based on this relationship, we proved a uniform bound on the estimation error under suitable parameter choices and an appropriate initial error condition. Simulation results on a linearized four-water-tank process demonstrated that the proposed OORD-MHE approach is able to attenuate the influence of online outliers.}

\section*{Declaration of Generative AI and AI-assisted technologies in the writing process}
 
During the revision of this work, the authors used ChatGPT to improve readability and clarity of the manuscript and to support consistency checks of the mathematical exposition. The authors have reviewed and edited all AI-assisted content as needed and take full responsibility for the content of the publication. All mathematical arguments were verified by the authors.


\begin{thebibliography}{99}

\bibitem{kalman1960new}
R.~E. Kalman.
\newblock A new approach to linear filtering and prediction problems.
\newblock {\em Transactions of the ASME--Journal of Basic Engineering}, 82(1):35--45, 1960.

\bibitem{mayne2014model}
D.~Q. Mayne.
\newblock Model predictive control: Recent developments and future promise.
\newblock {\em Automatica}, 50(12):2967--2986, 2014.

\bibitem{anderson2005optimal}
B.~D.~O. Anderson and J.~B. Moore.
\newblock Optimal filtering.
\newblock {\em Courier Corporation}, 2005.

\bibitem{rao2001constrained}
C.~V. Rao, J.~B. Rawlings, and  J.~H. Lee.
\newblock Constrained linear state estimation--a moving horizon approach.
\newblock {\em Automatica}, 37(10):1619--1628, 2001.

\bibitem{rawlings2020model}
J.~B. Rawlings, D.~Q. Mayne, and M.~M. Diehl.
\newblock Model predictive control: Theory, computation, and design. 2nd edition.
\newblock {\em Madison, WI: Nob Hill Publishing}, 2020.

\bibitem{rao2003constrained}
C.~V. Rao, J.~B. Rawlings, and  D.~Q. Mayne.
\newblock Constrained state estimation for nonlinear discrete-time systems: Stability and moving horizon approximations.
\newblock {\em IEEE Transactions on Automatic Control}, 48(2):246--258, 2003.

\bibitem{alessandri2008moving}
A. Alessandri, M. Baglietto, and G. Battistelli.
\newblock Moving-horizon state estimation for nonlinear discrete-time systems: New stability results and approximation schemes.
\newblock {\em Automatica}, 44(7):1753--1765, 2008.


\bibitem{zou2020mhe}
L. Zou, Z. Wang, J. Hu, and D. Zhou.
\newblock Moving horizon estimation with unknown inputs under dynamic quantization effects.
\newblock {\em IEEE Transactions on Automatic Control}, 65(12):5368--5375, 2020.

\bibitem{schiller2023lyapunov}
J.~D. Schiller, S. Muntwiler, J. K{\"o}hler, M.~N. Zeilinger, and M.~A. M{\"u}ller.
\newblock A Lyapunov function for robust stability of moving horizon estimation.
\newblock {\em IEEE Transactions on Automatic Control}, 68(12):7466--7481, 2023.

\bibitem{yin2021event}
X. Yin and J. Liu.
\newblock Event-triggered state estimation of linear systems using moving horizon estimation.
\newblock {\em IEEE Transactions on Control Systems Technology}, 29(2):901--909, 2021.

\bibitem{schneider2017partition}
R. Schneider.
\newblock A solution for the partitioning problem in partition-based moving-horizon estimation.
\newblock {\em IEEE Transactions on Automatic Control}, 62(6):3076--3082, 2017.

\bibitem{yan2025selftuning}
M. Yan, M. Han, A. W.-K. Law, and X. Yin.
\newblock Self-tuning moving horizon estimation of nonlinear systems via physics-informed machine learning Koopman modeling.
\newblock {\em AIChE Journal}, 71(2):e18649, 2025.



\bibitem{van2012subspace}
P.~Van Overschee and B.~De Moor.
\newblock Subspace identification for linear systems: Theory--Implementation--Applications.
\newblock {\em Springer Science \& Business Media}, 2012.



\bibitem{wolff2024robust}
T.~M. Wolff, V.~G. Lopez, and M.~A. M{\"u}ller.
\newblock Robust data-driven moving horizon estimation for linear discrete-time systems.
\newblock {\em IEEE Transactions on Automatic Control}, 69(8):5598--5604, 2024.



\bibitem{willems2005note}
J.~C. Willems, P. Rapisarda, I. Markovsky, and B.~L.~M. De~Moor.
\newblock A note on persistency of excitation.
\newblock {\em Systems \& Control Letters}, 54(4):325--329, 2005.

\bibitem{markovsky2008data}
I. Markovsky and P. Rapisarda.
\newblock Data-driven simulation and control.
\newblock {\em International Journal of Control}, 81(12):1946--1959, 2008.

\bibitem{coulson2019data}
J. Coulson, J. Lygeros, and F. D{\"o}rfler.
\newblock Data-enabled predictive control: In the shallows of the DeePC.
\newblock {\em European Control Conference}, 307--312, 2019.


\bibitem{liu2004line}
H. Liu, S. Shah, and W. Jiang.
\newblock On-line outlier detection and data cleaning.
\newblock {\em Computers \& Chemical Engineering}, 28(9):1635--1647, 2004.

\bibitem{chen2022gross}
S.-Y. Chen, Y.-W. Wang, and Y.-Q. Ni.
\newblock Gross outlier removal and fault data recovery for SHM data of dynamic responses by an annihilating filter-based Hankel-structured robust PCA method.
\newblock {\em Structural Control and Health Monitoring}, 29(12):e3144, 2022.

\bibitem{garces2011outliers}
H. Garces and D. Sbarbaro.
\newblock Outliers detection in environmental monitoring databases.
\newblock {\em Engineering Applications of Artificial Intelligence}, 24(2):341--349, 2011.


\bibitem{duran2024outlier}
G. Duran-Martin, M. Altamirano, A.~Y. Shestopaloff, L. S{\'a}nchez-Betancourt, J. Knoblauch, M. Jones, F.-X. Briol, and K. Murphy.
\newblock Outlier-robust Kalman filtering through generalised Bayes.
\newblock {\em International Conference on Machine Learning}, 235:12138--12171, 2024.

\bibitem{cao2025robust}
W. Cao, C. Liu, Z. Lan, S.~E. Li, W. Pan, and A. Alessandri.
\newblock Robust Bayesian inference for moving horizon estimation.
\newblock {\em Automatica}, 173:112108, 2025.

\bibitem{gibbs2013new}
R.~G. Gibbs.
\newblock New Kalman filter and smoother consistency tests.
\newblock {\em Automatica}, 49(10):3141--3144, 2013.

\bibitem{alessandri2016moving}
A. Alessandri and M. Awawdeh.
\newblock Moving-horizon estimation with guaranteed robustness for discrete-time linear systems and measurements subject to outliers.
\newblock {\em Automatica}, 67:85--93, 2016.


\bibitem{roth2013student}
M. Roth, E. {\"O}zkan, and F. Gustafsson.
\newblock A Student's t filter for heavy tailed process and measurement noise.
\newblock {\em International Conference on Acoustics, Speech and Signal Processing}, 5770--5774, 2013.

\bibitem{akkaya2008robust}
A.~D. Akkaya and M.~L. Tiku.
\newblock Robust estimation in multiple linear regression model with non-Gaussian noise.
\newblock {\em Automatica}, 44(2):407--417, 2008.

\bibitem{boncelet1983approach}
C.~G. Boncelet and B.~W. Dickinson.
\newblock An approach to robust Kalman filtering.
\newblock {\em IEEE Conference on Decision and Control}, 304--305, 1983.


\bibitem{gandhi2009robust}
M.~A. Gandhi and L. Mili.
\newblock Robust Kalman filter based on a generalized maximum-likelihood-type estimator.
\newblock {\em IEEE Transactions on Signal Processing}, 58(5):2509--2520, 2010.

\bibitem{liu2017maximum}
X. Liu, B. Chen, B. Xu, Z. Wu, and P. Honeine.
\newblock Maximum correntropy unscented filter.
\newblock {\em International Journal of Systems Science}, 48(8):1607--1615, 2017.

\bibitem{chen2017maximum}
B. Chen, X. Liu, H. Zhao, and J.~C. Principe.
\newblock Maximum correntropy Kalman filter.
\newblock {\em Automatica}, 76:70--77, 2017.




\bibitem{nicoletti2009computational}
M. do Carmo Nicoletti, L.~C. Jain, and  R.~C.~Giordano.
\newblock Computational intelligence techniques as tools for bioprocess modelling, optimization, supervision and control.
\newblock {\em Berlin/Heidelberg: Springer}, 2009.

\bibitem{bourgeois2001line}
W. Bourgeois, J.~E. Burgess, and  R.~M. Stuetz.
\newblock On-line monitoring of wastewater quality: a review.
\newblock {\em Journal of Chemical Technology \& Biotechnology}, 76(4):337--348, 2001.

\bibitem{candes2011robust}
E.~J. Cand{\`e}s, X. Li, Y. Ma, and J. Wright.
\newblock Robust principal component analysis?
\newblock {\em Journal of the ACM}, 58(3):1--37, 2011.


\bibitem{huber2011robust}
P.~J. Huber.
\newblock Robust statistics.
\newblock {\em International Encyclopedia of Statistical Science}, 1248--1251, 2011.

\bibitem{berberich2020trajectory}
J. Berberich and F. Allg{\"o}wer.
\newblock A trajectory-based framework for data-driven system analysis and control.
\newblock {\em European Control Conference}, 1365--1370, 2020.






\bibitem{knufer2020time}
S. Kn{\"u}fer and M.~A. M{\"u}ller.
\newblock Time-discounted incremental input/output-to-state stability.
\newblock {\em IEEE Conference on Decision and Control}, 5394--5400, 2020.

\bibitem{dorflerref}
F. D{\"o}rfler, J. Coulson, and I. Markovsky.
\newblock Bridging direct and indirect data-driven control formulations via regularizations and relaxations.
\newblock {\em IEEE Transactions on Automatic Control}, 68(2):883--897, 2023.






\bibitem{jensen1906fonctions}
J.~L.~W.~V. Jensen.
\newblock Sur les fonctions convexes et les in{\'e}galit{\'e}s entre les valeurs moyennes.
\newblock {\em Acta Mathematica}, 30(1):175--193, 1906.

\bibitem{andersson2019casadi}
J.~A.~E. Andersson, J. Gillis, G. Horn, J.~B. Rawlings, and M. Diehl.
\newblock {CasADi}: A software framework for nonlinear optimization and optimal control.
\newblock {\em Mathematical Programming Computation}, 11(1):1--36, 2019.

\bibitem{wachter2006implementation}
A. W{\"a}chter and L.~T. Biegler.
\newblock On the implementation of an interior-point filter line-search algorithm for large-scale nonlinear programming.
\newblock {\em Mathematical Programming}, 106(1):25--57, 2006.

\end{thebibliography}



\appendix
\section*{Appendix}
\section{Properties of the truncated-quadratic loss}   
\label{appendix1}
\begin{prop}
    \label{propostion-ell}
    For any $\sigma, \delta \in \mathbb{R}^{n_y}$, the following inequalities hold for $\ell$ in \eqref{outliereq-loss-sigma-y}:
    \begin{align}
        \ell(\sigma, \bar{l}) &\leq \| \sigma \|^2 \label{outliereq-ell1}\\
        \sqrt{\ell(\sigma + \delta, \bar{l})} &\leq \sqrt{\ell(\sigma, \bar{l})} + \| \delta \| \label{outliereq-ell2} \\
        \ell(\sigma + \delta, \bar{l}) & \leq 2\ell(\sigma, \bar{l}) + 2\ell(\delta, \bar{l})\label{outliereq-ell3} 
    \end{align}
\end{prop}
\begin{pf}
    \eqref{outliereq-ell1} is ensured by the definition of $\ell$ in \eqref{outliereq-loss-sigma-y}. For \eqref{outliereq-ell2}, by the triangle inequality, we have
    \[
    \begin{aligned}
        & \sqrt{\ell(\sigma + \delta, \bar{l})} \leq \min \left\{ \| \sigma \| + \| \delta \|, \sqrt{\bar{l}} \right\} \\
        \leq~ &\min \left\{ \| \sigma \| , \sqrt{\bar{l}} \right\} + \| \delta \|
        = \sqrt{\ell(\sigma, \bar{l})} + \| \delta \|{.}
    \end{aligned}
    \]
    For \eqref{outliereq-ell3}, by $\|a+b\|^2\leq 2\|a\|^2+2\|b\|^2$, we obtain
    \[
    \begin{aligned}        
        & \ell(\sigma + \delta, \bar{l}) \leq \min \left\{ 2\| \sigma \|^2 + 2\| \delta \|^2, \bar{l} \right\} \\
        \leq ~ & \min \left\{ 2\| \sigma \|^2 , \bar{l} \right\} + \min \left\{ 2\| \delta \|^2 , \bar{l} \right\} 
        \\
        \leq ~& 2 \min \left\{ \| \sigma \|^2 , \bar{l} \right\} + 2 \min \left\{ \| \delta \|^2 , \bar{l} \right\} 
        \leq 2\ell(\sigma, \bar{l}) + 2\ell(\delta, \bar{l}){.}
    \end{aligned}
    \]
    This ends the proof. \qed
\end{pf}

\section{Proof of Proposition 1}       
\label{appendix2}
{

For notational simplicity, we use
$(g_k,\hat{\bs{x}}_k)$ to denote the solution $(g_k,\hat{\bs{x}}_k, H_{{L\ominus k}+1}(\tilde{x}^{{\color{blue}\off}}_{[0,N^{{\color{blue}\off}}]})g_k-\hat{\bs{x}}_k, \tilde{y}_{[{l_k},k-1]} - H_{{L\ominus k}}(\tilde{y}^{{\color{blue}\off}}_{[0,N^{{\color{blue}\off}}-1]})g_k )$ in the
subsequent analysis as discussed in Remark~\ref{remark:solution}.
}

\noindent\textbf{Step I}: We use $J(g_k,\hat{\bs{x}}_k)$ to denote $J_k$ in \eqref{outliereq-loss-OORD-MHE}. Following the proof of Theorem 2 in \cite{wolff2024robust}, we construct a candidate solution $(g^{\true}_k, x_{[l_k, k]})$ for \eqref{outliereq-OORD-MHE} where 
\begin{equation}
    \label{outliereq-g-true}
    g^{\true}_k \coloneqq
    \begin{bmatrix} 
        H_{{L\ominus k}}\left(u^{{\color{blue}\off}}_{[0,N^{{\color{blue}\off}}-1]}\right) \\ 
        H_{1}\left(x^{{\color{blue}\off}}_{[0,N^{{\color{blue}\off}}-{L\ominus k}]} \right)
    \end{bmatrix}^\dagger   
    \begin{bmatrix} 
        u_{[{l_k},k-1]} \\ 
        x_{{l_k}}
    \end{bmatrix}{.}
\end{equation}  
By Assumptions~\ref{assumption-controllable} and \ref{assumption-PE}, the concatenated matrix has full row rank.

Therefore, Assumption~\ref{assumption-trajectory-bound} and 
{$\bar g\geq G(L)$} imply that $\| g^{\true}_k \| \leq \bar{g}$. Further, $x_{[{l_k},k]}$ satisfies the state constraint $\mathbb X$. As a result, $(g^{\true}_k,~x_{[{l_k},k]})$ is feasible for OORD-MHE.

\noindent \textbf{Step II}: 
By Lemma~\ref{lemma-hankel} and \eqref{outliereq-willems-x}, \eqref{outliereq-tilde-y}, \eqref{outliereq-OORD-MHE-hankel}, \eqref{outliereq-g-true}, we have
\begin{subequations}
    \label{outliereq-trues}
    \begin{align}
        & y_{j} = H_1(y^{{\color{blue}\off}}_{[{j-l_k},j-k+N^{\color{blue}\off}]})g^{\true}_k \label{outliereq-true-y}\\
        & x_{j} = H_1(x^{{\color{blue}\off}}_{[{j-l_k},j-k+N^{\color{blue}\off}]})g^{\true}_k \label{outliereq-true-x} \\
        & \sigma^{y, \true}_{j|k} \!=\! \epsilon_j+o_j - H_{1}(\epsilon^{{\color{blue}\off}}_{y,[{j-l_k},j-k+N^{{\color{blue}\off}}]})g^{\true}_k  \label{outliereq-sigma-y-true} \\
         &\sigma^{x, \true}_{j|k} \!=\! 
        H_{1}(\epsilon^{{\color{blue}\off}}_{x,[{j-l_k},j-k+N^{{\color{blue}\off}}]})g^{\true}_k \label{outliereq-sigma-x-true}
    \end{align}
\end{subequations}
where $j\in \mathbb{Z}_{[{l_k},k-1]}$ for $y_j$ and $\sigma^{y, \true}_{j|k}$; $j\in \mathbb{Z}_{[{l_k},k]}$ for $x_j$ and $\sigma^{x, \true}_{j|k}$. 
Note that since $\bar{H}_\epsilon = \sqrt{N^{{\color{blue}\off}}}\bar{\epsilon}^{{\color{blue}\off}}$ and the offline noise sequences are bounded, we have
\begin{subequations}
    \begin{align}
    \| H_1(\epsilon^{{\color{blue}\off}}_{y,[{j-l_k},j-k+N^{{\color{blue}\off}}]}) \|_F & \leq \bar{H}_\epsilon \label{outliereq-H-max1}{,}\\
    \| H_1(\epsilon^{{\color{blue}\off}}_{x,[{j-l_k},j-k+N^{{\color{blue}\off}}]}) \|_F & \leq \bar{H}_\epsilon  \label{outliereq-H-max2}{.}
    \end{align}
\end{subequations}
It follows that, by \eqref{outliereq-sigma-x-true} and \eqref{outliereq-H-max2},
\begin{align}
    \label{outliereq-sigma-x-true-bound}
    \| \sigma^{x, \true}_{j|k} \| \leq
        \bar{H}_\epsilon \bar g
\end{align}
and by \eqref{outliereq-sigma-y-true}, \eqref{outliereq-ell1}, \eqref{outliereq-ell3}, and \eqref{outliereq-H-max1}, 
\begin{align}
    \label{outliereq-l-true-bound}
        {\color{blue}\ell}(\sigma^{y, \true}_{j|k}, \bar{l}) \leq
        2{\color{blue}\ell}(\epsilon_j+o_j, \bar{l}) + 2\bar{H}_\epsilon^2 \bar g^2{.}
\end{align}
Then by \eqref{outliereq-prior-error}, \eqref{outliereq-sigma-x-true}, \eqref{outliereq-sigma-x-true-bound}, and \eqref{outliereq-l-true-bound}, $J(g^{\true}_k, x_{[{l_k},k]})$ can be bounded as
\begin{align}
    \label{outliereq-J-true-bound}
        & J(g^{\true}_k, x_{[{l_k},k]}) \leq~  {\bar \lambda (P)}\rho^L \bar{s}^2   + 2\sum_{j={l_k}}^{k-1}{\color{blue}\ell}(\epsilon_j+o_j, \bar{l})\notag\\
        &\quad\quad+ ({\color{blue}\mu_x}(L+1)+{\color{blue}\mu_g}+2L)\bar{H}_\epsilon^2 \bar g^2.
\end{align}

\noindent\textbf{Step III}: {\color{blue}Using \eqref{outliereq-define-gamma-1} and \eqref{outliereq-bar-l}, we bound the output penalty as
\[
\sum_{j=l_k}^{k-1}\ell(\epsilon_j+o_j,\bar l)\leq
L\bar\epsilon^2+|\Gamma_k^o|a.
\]
Then, combining the above inequality with $a=\alpha_1(L,\bar H_\epsilon\bar g,\bar s)$, \eqref{alpha1}, and \eqref{outliereq-J-true-bound} yields
\begin{align}
\label{outliereq-J-true-bound1}
    J(g^{\true}_k, x_{[{l_k},k]}) & \leq a - (2L+1)\bar \epsilon^2+2(L\bar\epsilon^2 + |\Gamma_k^o|a) \notag \\
    &<
    (2|\Gamma_k^o|+1)\alpha_1(L, \bar{H}_\epsilon\bar g, \bar{s}){.}
\end{align}}
Then, {\color{blue}by} global optimality of $(g^*_k,\hat x^*_{[{l_k},k]|k})$, we have
{\begin{align}
    \label{outliereq-conflict-1}
        & \qquad\sum_{j={l_k}}^{k-1}  {\color{blue}\ell}( \sigma^{y, *}_{j|k}, \bar{l})
        \leq  J(g^*_k, \hat{\bs{x}}_k^*) \leq
        J(g^{\true}_k, x_{[{l_k},k]})\notag\\
        &\stackrel{\eqref{outliereq-J-true-bound1}}{<}
    (2|\Gamma_k^o|+1)\alpha_1(L, \bar{H}_\epsilon{\bar g}, \bar{s}){.}
\end{align}}
Since $a = \alpha_1(L, \bar{H}_\epsilon{\bar g}, \bar{s})$, if $|\Gamma^*_{k, a}| \geq 2|\Gamma_k^o|+1$, then\vspace{-10pt}
\[
\begin{aligned}
     &\sum_{j={l_k}}^{k-1} {\color{blue}\ell}( \sigma^{y, *}_{j|k}, \bar{l}) 
    \stackrel{\eqref{outliereq-loss-sigma-y}}{\geq} \sum_{j\in \Gamma^*_{k, a}}
    \min\{ \|\sigma^{y, *}_{j|k}\|^2, \bar{l}\} \\
    &\stackrel{\eqref{outliereq-define-gamma-2}\eqref{outliereq-bar-l}}{\geq}  ~~ \sum_{j\in \Gamma^*_{k, a}} \alpha_1(L, \bar{H}_\epsilon{\bar g}, \bar{s}) 
    \geq~    (2|\Gamma_k^o|+1)  \alpha_1(L, \bar{H}_\epsilon{\bar g}, \bar{s})
\end{aligned}
\]
which contradicts \eqref{outliereq-conflict-1} due to $\bar\epsilon>0$. So, $|\Gamma^*_{k, a}| < 2|\Gamma_k^o|+1$. Since $|\Gamma^*_{k, a}|$ is an integer, we have $|\Gamma^*_{k, a}| \leq 2|\Gamma_k^o|$,
which completes the proof. \qed

\section{Proof of Theorem 1}  
\label{appendix3}

\noindent 
For the optimal solution, we have\vspace{-10pt}
\begin{subequations}
    \label{outliereq-optimals}
    \begin{align}
        y_{j|k}^{*} & = H_1(y^{{\color{blue}\off}}_{[{j-l_k},j-k+N^{\color{blue}\off}]})g^{*}_k \label{outliereq-optimal-y}{,}\\
        x_{j|k}^{*} & = H_1(x^{{\color{blue}\off}}_{[{j-l_k},j-k+N^{\color{blue}\off}]})g^{*}_k{,}\label{outliereq-optimal-x} \\
        \sigma^{y, *}_{j|k} & = \tilde{y}_j - y_{j|k}^{*} - H_{1}(\epsilon^{{\color{blue}\off}}_{y,[{j-l_k},j-k+N^{{\color{blue}\off}}]})g^{*}_k{,}  \label{outliereq-optimal-sigma-y} \\
        \sigma^{x, *}_{j|k} & = x_{j|k}^{*} - \hat x^*_{j|k} +
        H_{1}(\epsilon^{{\color{blue}\off}}_{x,[{j-l_k},j-k+N^{{\color{blue}\off}}]})g^{*}_k{.} \label{outliereq-optimal-sigma-x}
    \end{align}
\end{subequations}
For notational convenience, define $\Delta^x_j \coloneqq \| x_j - x_{j|k}^{*} \|$ and  $\Delta^y_j \coloneqq \| y_j - y_{j|k}^{*} \|$.  
Following \cite{wolff2024robust}, by Assumption \ref{assumption-detectable}, e-UOSS~\eqref{outliereq-e-uoss}, $\bar{x}_k = \hat x^*_{k|k}$, and \eqref{outliereq-optimals}, $\| x_k-\bar{x}_k\|$ satisfies\vspace{-10pt}
    \begin{align}
    \label{part1}
        & \| x_k-\bar{x}_k\| \stackrel{\eqref{outliereq-e-uoss}\eqref{outliereq-H-max2}\eqref{outliereq-optimal-sigma-x}}{\leq} 
            p_0\eta^{{L\ominus k}} \Delta^x_{{l_k}}   \notag \\
            & 
            + \sum_{j={l_k}}^{k-1} p_1 \eta^{k-j-1} \Delta^y_j
            + \| \sigma^{x, *}_{k|k} \| 
            + \bar{H}_\epsilon \| g^*_k \| {.}
    \end{align}
\noindent\textbf{Step I}: 
\noindent If $j\in\bar{\Gamma}_{k, a}$, we have\vspace{-10pt}
{\color{blue}\begin{align}
    \label{delta-y-small}
        \Delta^y_j & ~~~~~~~~\leq ~~~~~~~~\| y_j - \tilde y_j\| +\|\tilde y_j - y_{j|k}^{*} \|\notag\\
        & \stackrel{\eqref{outliereq-tilde-y}\eqref{outliereq-optimal-sigma-y}\eqref{outliereq-H-max1}}{\leq} \|\epsilon_j+o_j\| + \|\sigma_{j|k}^{y, *}\|+\bar H_\epsilon\|g_k^*\|\notag \\
        &\stackrel{~~\eqref{outliereq-g-max}\eqref{outliereq-define-gamma-1}\eqref{outliereq-define-gamma-2}~~}{\leq} \bar \epsilon + \sqrt{\alpha_1(L,\bar H_\epsilon\bar g,\bar s)} + \bar H_\epsilon\bar g \notag\\
         & ~~~~~~~~\leq~~~~~~~~ c_1(L, \bar{H}_\epsilon{\bar g}, \bar{s})
    \end{align}}
where {$c_1(L, z, s) \coloneqq (\sqrt{2L{+1}}+1)\bar{\epsilon} + z + \sqrt{{\bar \lambda (P)}\rho^L}s + \alpha_2(L, z)$ with $\alpha_2(L, z) \coloneqq \sqrt{{\color{blue}\mu_x}(L+1)+{\color{blue}\mu_g}+2L }z$.}
If $j\in \Gamma_{k, a}$, denote $\Gamma_{k, a} = \{j_1, \cdots, j_{|\Gamma_{k, a}|}\}$, $j_1 <  \cdots < j_{|\Gamma_{k, a}|}$.
For $j_1\in\Gamma_{k, a}$, which is the smallest time instant within $\Gamma_{k, a}$, it holds that\vspace{-10pt}
\begin{align}
    \Delta^x_{j_1} 
    \stackrel{\eqref{outliereq-e-uoss}\eqref{delta-y-small}}{\leq } 
        p_0\eta^{j_1-l_k} \Delta^x_{{l_k}} 
        + \frac{1-\eta^{j_1-l_k}}{1-\eta} p_1 c_1(L, \bar{H}_\epsilon{\bar g}, \bar{s}) {.}
\end{align}
And for each $j_i\in \Gamma_{k, a}$ with $i\neq 1$, applying a similar argument yields\vspace{-10pt}
    \begin{align}
    \label{outliereq-delta-x-ji}
        \Delta^x_{j_i}
        \stackrel{\eqref{outliereq-e-uoss}\eqref{delta-y-small}}{\leq }& 
            p_0\eta^{j_i-l_k} \Delta^x_{{l_k}} + p_1 \sum_{m=1}^{i-1} \eta^{j_i-j_m-1} \Delta^y_{j_m}
             \notag \\
            & + \frac{1-\eta^{j_i-l_k}}{1-\eta} p_1 c_1(L, \bar{H}_\epsilon{\bar g}, \bar{s})
    \end{align}
{Since $y(k|x_0, u_{[0,k]}) = C x(k|x_0,u_{[0,k]}) + D u_k$, there exists a constant $p_2\geq {\|C\|_2}$, such that 
\begin{align}
    \label{outliereq-y-discrepancy}
    &\quad\| y(k|x_0, u_{[0,k]}) - y(k|x_0', u_{[0,k]}) \| 
      \notag\\ 
    & \leq{p_2}\| x(k|x_0, u_{[0,k]}) - x(k|x_0', u_{[0,k]})\|{.}
\end{align}}
Therefore, we have
\begin{align}
\label{outliereq: delta-y}
    & \eta^{k-j_i-1}\Delta^y_{j_i} \notag\\ 
    \stackrel{\eqref{outliereq-y-discrepancy}\eqref{outliereq-delta-x-ji}}{\leq} & 
    b_1 + b_2\eta^{k-j_i-1}+ b_3\sum_{m=1}^{i-1} \eta^{k-j_m-1} \Delta^y_{j_m}
\end{align}
where
$b_1 = p_2\eta^{{L\ominus k}-1}
        p_0 \Delta^x_{{l_k}} $,
$b_2 = \dfrac{b_3\eta c_1(L, \bar{H}_\epsilon{\bar g}, \bar{s})}{1-\eta}$,
and $b_3 = \dfrac{p_1p_2}{\eta}$. For any $i \in \mathbb{Z}_{[2,|\Gamma_{k, a}|]}$, let $h_{j_1} \coloneqq b_1 + b_2\eta^{k-j_1-1}$ and $h_{j_i} \coloneqq b_1 + b_2\eta^{k-j_i-1} + b_3\sum_{m=1}^{i-1}h_{j_m}$.
So, \eqref{outliereq: delta-y} leads to\vspace{-10pt}
\begin{equation}
    \label{delta-h} 
    \sum_{j\in\Gamma_{k, a}} \eta^{k-j-1}\Delta^y_j \leq \sum_{j\in\Gamma_{k, a}} h_j{.}
\end{equation}
We now prove that the right-hand side of \eqref{delta-h} is maximized when $\Gamma_{k, a}$ consists of the  $|\Gamma_{k, a}|$ most recent time indices, i.e., $\Gamma_{k, a}=\{k-|\Gamma_{k, a}|, \cdots, k-1\}$.
If $|\Gamma_{k, a}|=0$, the claim is immediate. 
Let $|\Gamma_{k, a}|\geq1$ and write $\Gamma_{k,a}=\{j_1,\ldots,j_{|\Gamma_{k, a}|}\}$ with $j_1<\cdots<j_{|\Gamma_{k, a}|}$. Since $j_i\leq k-|\Gamma_{k, a}|-1+i$, $\eta^{k-j_i-1}\leq\eta^{|\Gamma_{k, a}|-i}$.
Moreover, the recursion defining $h_{j_i}$ is nondecreasing in each of the weights $\eta^{k-j_m-1}$. Hence, by induction, $h_{j_i}\leq h_i'$, where $h_0'=0$, $h_i' = b_1+b_2\eta^{|\Gamma_{k, a}|-i} + b_3\sum_{m=1}^{i-1}h_m'$.
It yields that\vspace{-10pt}
\begin{equation}
    \label{h-h'}
    \sum_{j\in\Gamma_{k, a}} h_j \leq \sum_{i=0}^{|\Gamma_{k, a}|} h'_{i}{.}
\end{equation}
We now calculate the analytical expression of the right-hand side of \eqref{h-h'}. 
For $i\in \mathbb{Z}_{[0,|\Gamma_{k, a}|]}$, define $S_{i} \coloneqq \sum_{m=0}^i h'_{m}$. Then, it follows that
$
    S_{1} = b_1 + b_2 \eta^{|\Gamma_{k, a}| - 1},~ 
    S_{i}-S_{i-1} = b_1 + b_2 \eta^{|\Gamma_{k, a}| - i} + b_3 S_{i-1},~i\in \mathbb{Z}_{[1,|\Gamma_{k, a}|]}
$.
Therefore, \vspace{-10pt}
\begin{equation}
    \label{outliereq-Si}
    \begin{aligned}
        S_{i} 
            & = b_1 \frac{(1+b_3)^i-1}{b_3} + b_2\eta^{|\Gamma_{k, a}| - i} \beta_0(i)
    \end{aligned}
\end{equation}
where $\beta_0(0)=0, ~\forall i\in\mathbb{Z}_{\geq 1}, ~\beta_0(i) = \sum_{m=0}^{i-1} (1+b_3)^m\eta^m$. Combining \eqref{delta-h} and \eqref{h-h'}, we can conclude  that $
p_1\sum_{j\in\Gamma_{k, a}} \eta^{k-j-1}\Delta^y_j 
\stackrel{}{\leq} 
p_1S_{|\Gamma_{k, a}|}$. 
As a result, we have 
    \begin{align}
    \label{delta-y-sum}
        p_1\sum_{j\in\Gamma_{k, a}} \eta^{k-j-1}\Delta^y_j
        &\stackrel{\eqref{outliereq-Si}}{\leq}  
            (\beta_1(|\Gamma_{k, a}|)-p_0)\eta^{{L\ominus k}}\Delta^x_{{l_k}} \notag \\ 
            &+ c_1(L, \bar{H}_\epsilon{\bar g}, \bar{s})\beta_2(|\Gamma_{k, a}|)
    \end{align}
where ${\beta_1}(i)=p_0 (\dfrac{p_1p_2}{\eta}+1)^i$ and ${\beta_2}(i)=\dfrac{p_1^2p_2 \beta_0(i)}{1-\eta}$. 

\noindent\textbf{Step II}: 
\noindent Applying the result in Step I to \eqref{part1} yields 
\begin{align}
    \label{step4}
        & \| x_k-\bar{x}_k\| 
        \stackrel{\eqref{delta-y-small} \eqref{delta-y-sum}\eqref{outliereq-optimal-sigma-x}\eqref{outliereq-H-max2}}{\leq}            
            \beta_1(|\Gamma_{k, a}|)\eta^{{L\ominus k}}   \notag \\
           &  \bigg[ \| x_{{l_k}}-\bar{x}_{{l_k}}\|+ \| \bar{x}_{{l_k}}-\hat{x}^*_{{l_k}|k}\|
            + \bar{H}_\epsilon\| g^*_k \|  
            + \| {\sigma}^{x, *}_{{l_k}|k} \| \bigg]\notag \\
           & +   \| {\sigma}^{x, *}_{k|k} \| 
            + \bar{H}_\epsilon \| g^*_k \| 
             + c_1(L, \bar{H}_\epsilon{\bar g}, \bar{s}) \beta_2(|\Gamma_{k, a}|) +  \notag \\
           &  \sum_{j\in\bar{\Gamma}_{k, a}} p_1 \eta^{k-j-1} (\| \epsilon_j+o_j \| + \| {\sigma}^{y, *}_{j|k}\| + \bar{H}_\epsilon\| g^*_k\|  )
    \end{align}
{Further, we denote
\[
\begin{aligned}
    q(|\Gamma_{k, a}|) = \max \bigg\{\frac{\beta_1(|\Gamma_{k, a}|)}{{ {\underline{\lambda}(P)}}},
    \frac{\beta_1(|\Gamma_{k, a}|)\eta+1}{{\color{blue}\mu_x}}, p_1, \\
    \frac{\beta_1(|\Gamma_{k, a}|)\eta+1}{{\color{blue}\mu_g}} +\frac{p_1}{{\color{blue}\mu_g}(1-\eta)} \bigg\}
\end{aligned}
\]
Then, by combining $\eta<\rho$ and \eqref{step4}, we have
    \begin{align}
    \label{step5}
        &  \| x_k-\bar{x}_k\| {~\leq~}  {q(|\Gamma_{k, a}|) {\underline{\lambda} (P)}}\rho^{{L\ominus k}}\| x_{{l_k}}-\bar{x}_{{l_k}}\| + { q(|\Gamma_{k, a}|)} \tilde{J}^* \notag \\
        &  + c_1(L, \bar{H}_\epsilon{\bar g}, \bar{s}) \beta_2(|\Gamma_{k, a}|)
            + \sum_{j\in\bar{\Gamma}_{k, a}} p_1 \| \epsilon_j+o_j \|              
    \end{align}}
where
\begin{align}
\label{outliereq-J*}
    \tilde{J}^* = & { \underline{\lambda} (P)}{\rho}^{{L\ominus k}}\| \bar{x}_{{l_k}}-\hat{x}^*_{{l_k}|k}\| 
            + {\color{blue}\mu_x} \| {\sigma}^{x, *}_{[{l_k},k]|k} \| \notag \\
            & + {\color{blue}\mu_g} \bar{H}_\epsilon\| g^*_k\|  
            + \sum_{j\in\bar{\Gamma}_{k, a}} \| {\sigma}^{y, *}_{j|k}\|{.}
\end{align}    
\noindent\textbf{Step III}: 
\noindent This part is based on the proof of Theorem 2 in \cite{wolff2024robust}.
By \eqref{outliereq-define-gamma-2}, \eqref{outliereq-gamma-derive}, and \eqref{outliereq-bar-l}, we have $\forall j\in \bar{\Gamma}_{k, a}$, 
    $\sum_{j\in\bar{\Gamma}_{k, a}}\ell({\sigma}^{y, *}_{j|k}, \bar{l}) 
    = \sum_{j\in\bar{\Gamma}_{k, a}}\| {\sigma}^{y, *}_{j|k} \|^2$.
Applying this relationship to the optimal loss $J(g^*_k, \hat{x}^*_{[k- {L\ominus k},k]|k})$ yields\vspace{-10pt} 
\begin{align}
    \label{outliereq-pre1}
             & J(g^*_k, \hat{x}^*_{[k- {L\ominus k},k]|k})  \geq
                 \rho^{{L\ominus k}} \| \hat{x}^*_{{l_k}|k} - \bar{x}_{{l_k}} \|^2_P
                 + {\color{blue}\mu_x} \| {\sigma}^{x, *}_{[{l_k},k]|k} \|^2\notag \\
                & \quad 
                + {\color{blue}\mu_g}\bar{H}^2_\epsilon \| g^*_k \|^2 
                + \sum_{j\in\bar{\Gamma}_{k, a}} \| {\sigma}^{y, *}_{j|k} \|^2  {.}
\end{align}
By applying Jensen's inequality \cite{jensen1906fonctions} to \eqref{outliereq-pre1}, it can be derived that
\begin{equation}
    \label{outliereq-left}
    \begin{aligned}
        & \alpha_3(L)\sqrt{J(g^*_k, \hat{x}^*_{[k- {L\ominus k},k]|k})} \geq
        \tilde{J}^*
    \end{aligned}
\end{equation}
where $\alpha_3(L) \coloneqq \sqrt{{ {\underline{\lambda} (P)}}\rho+{\color{blue}\mu_x}+ {\color{blue}\mu_g}+L}$. 
Using $\|a+b\|^2 \leq 2\|a\|^2+2\|b\|^2$, we can obtain
\begin{align}
    \label{outliereq-loss-true-bound}
    & J(g^{\true}_k, x_{[k- {L\ominus k},k]}) 
    \stackrel{\eqref{outliereq-g-max}\eqref{outliereq-ell1}\eqref{outliereq-sigma-y-true}\eqref{outliereq-H-max1}}{\leq} \sum_{j\in \Gamma_{k, a}}  \bar{l} + \notag \\
        &  \quad { {\bar \lambda (P)}\rho^{L\ominus k} \| x_{{l_k}} - \bar{x}_{{l_k}} \|^2}
        + {\color{blue}\mu_x} \| {\sigma}^{x, \true}_{[{l_k}, k]|k} \|^2 
         +  {\color{blue}\mu_g}\bar{H}^2_\epsilon \| g^{\true}_k \|^2\notag
        \\
        & \quad 
        + \sum_{j\in\bar{\Gamma}_{k, a}}  (2\|\epsilon_j+o_j \|^2 + 2\bar{H}_\epsilon^2\bar{g}^2){.}
\end{align}
Applying $\sqrt{a+b}\leq \sqrt{a}+\sqrt{b},~\forall a,b\geq 0$ to \eqref{outliereq-loss-true-bound} yields
    \begin{align}
    \label{outliereq-right}
        & \sqrt{J(g^{\true}_k, x_{[k- {L\ominus k},k]})} 
       ~\leq~  
        \sqrt{{ {\bar \lambda (P)}}\rho^{L\ominus k}} \| x_{{l_k}}-\bar{x}_{{l_k}} \| \notag \\
        & \quad \quad + \sqrt{{\color{blue}\mu_x}} \| {\sigma}^{x, \true}_{[{l_k}, k]|k} \| 
        + \sqrt{{\color{blue}\mu_g}}\bar{H}_\epsilon \| g^{\true}_k\| + \sum_{j\in\Gamma_{k, a}} \sqrt{\bar{l}}  \notag \\
        & \quad \quad + \sqrt{2}\sum_{j\in\bar{\Gamma}_{k, a}} (\| \epsilon_j+o_j\|+ \bar{H}_\epsilon\bar{g}){.}
    \end{align}
Then, combining~\eqref{outliereq-g-max}, \eqref{outliereq-prior-error}, \eqref{outliereq-bar-l}, \eqref{outliereq-sigma-x-true-bound}, \eqref{outliereq-left}, \eqref{outliereq-right}, and $J(g^*_k, \hat{x}^*_{[k- {L\ominus k},k]|k})  \leq J(g^{\true}_k, x_{[k- {L\ominus k},k]})$, we have 
\begin{align}
    \label{outliereq-part1}
        &\tilde{J}^* 
        { ~\leq~ } \notag\\ 
        &\alpha_3(L)\bigg[(|\Gamma_{k, a}|+1)\sqrt{{ {\bar \lambda (P)}}\rho^L}\bar{s} 
        + \sqrt{2}\sum_{j\in\bar{\Gamma}_{k, a}} (\| \epsilon_j+o_j\|)\bigg] \notag \\
        & + \alpha_4(L, \bar{H}_\epsilon{\bar g}) 
        + |\Gamma_{k, a}|\alpha_5(L, \bar{H}_\epsilon{\bar g})
    \end{align}
where $\alpha_4(L, \bar{H}_\epsilon {\bar g}) \coloneqq \alpha_3(L)\bar{H}_\epsilon\bar{g} (
    \sqrt{{\color{blue}\mu_x}(L+1)} + \sqrt{{\color{blue}\mu_g}} + \sqrt{2}L
)$ and $\alpha_5(L, \bar{H}_\epsilon{\bar g}) \coloneqq \alpha_3(L)(
    \sqrt{2L+{2}}\bar{\epsilon} + \alpha_2(L, \bar{H}_\epsilon{\bar g}) - \sqrt{2}\bar{H}_\epsilon\bar{g}
)$.

\noindent\textbf{Step IV}: 
\noindent Applying \eqref{outliereq-prior-error} and \eqref{outliereq-part1} to \eqref{step5} yields
\[
    \begin{aligned}
        &\| x_k-\bar{x}_k\| 
        {~\leq~}             
             \gamma_1(L, |\Gamma_{k, a}|) \sqrt{\rho}^L\bar{s} \\
            &\quad+ \gamma_2(L, |\Gamma_{k, a}|, \bar{H}_\epsilon\bar g)
            + \gamma_3(L, |\Gamma_{k, a}|)\sum_{j\in\bar{\Gamma}_{k, a}} \| \epsilon_j+o_j \| 
    \end{aligned}
\]
which is \eqref{outliereq-one-step},
where $\gamma_1(L, |\Gamma_{k, a}|)  = \beta_2(|\Gamma_{k, a}|)\sqrt{{ {\bar \lambda (P)}}} 
 +q(|\Gamma_{k, a}|)\alpha_3(L)(|\Gamma_{k, a}|+1)\sqrt{{ {\bar \lambda (P)}}} 
+ q(|\Gamma_{k, a}|){\underline{\lambda} (P)}
$, 
$\gamma_2(L, |\Gamma_{k, a}|, \bar{H}_\epsilon{\bar g})  = \beta_2(|\Gamma_{k, a}|)((\sqrt{2L+1}+1)\bar{\epsilon} + \bar{H}_\epsilon\bar{g} + \alpha_2(L, \bar{H}_\epsilon{\bar g})) + q(|\Gamma_{k, a}|)(\alpha_4(L, \bar{H}_\epsilon{\bar g}) + |\Gamma_{k, a}|\alpha_5(L, \bar{H}_\epsilon{\bar g})) $, $\gamma_3(L, |\Gamma_{k, a}|)  = p_1+\sqrt{2}q(|\Gamma_{k, a}|)\alpha_3(L)$.
\qed

\section{Proof of { Corollary 1}}  
\label{appendix4}
We prove this {corollary} by strong induction. Equation \eqref{outliereq-uniform-bound} holds at $k=0$ since $\| x_{0} - \bar{x}_{0}\| \leq \sqrt{\rho}^{L}\bar{s}$. 
Let $k\in\mathbb{Z}_{\geq 1}$ and suppose that
\[
\|x_t-\bar x_t\|\leq\bar s,
\qquad
\forall t\in\mathbb Z_{[0,k-1]}.
\]
If $k\leq L$, then $l_k=0$ and
\[
\|x_{l_k}-\bar x_{l_k}\|
\leq
\sqrt{\rho}^{L}\bar s
\leq
\sqrt{\rho}^{L-L\ominus k}\bar s.
\]
If $k>L$, then $l_k=k-L<k$.
Hence, by the induction hypothesis,
\[
\|x_{l_k}-\bar x_{l_k}\|
\leq
\bar s
=
\sqrt{\rho}^{L-L\ominus k}\bar s.
\]
Thus, \eqref{outliereq-prior-error} holds in both cases. Then by Theorem~\ref{theorem-one-horizon}, $\|x_k-\bar{x}_k\|$ is bounded by \eqref{outliereq-one-step}. Moreover, since
$\sum_{j\in \bar{\Gamma}_{k, a}} \| \epsilon_j+o_j \| \leq L \bar{\epsilon}$,
we can conclude that
\begin{align}
    & \| x_k-\bar{x}_k\| \stackrel{\eqref{outliereq-one-step}}{\leq}  \gamma_1(L, |\Gamma_{k, a}|) \sqrt{\rho}^L\bar{s} \notag \\
    & \quad\quad\quad + \gamma_2(L, |\Gamma_{k, a}|, \bar{H}_\epsilon \bar g)+\gamma_3(L, |\Gamma_{k, a}|) L \bar{\epsilon}{.}
\end{align}

Assumption \ref{assumption-outlier-number} implies that $|\Gamma_k^o|\leq n_o$.
By Proposition~\ref{prop-number}, we have $|\Gamma_{k, a}|\leq 3 |\Gamma_k^o| \leq 3n_o$. Since $\gamma_1$, $\gamma_2$, and $\gamma_3$ are all nondecreasing with respect to $|\Gamma_{k, a}|$, we have 
\begin{align}
    \label{outliereq-bound-k}
    & \| x_k-\bar{x}_k\| \leq \gamma_1(L, 3n_o) \sqrt{\rho}^L\bar{s} \notag \\
    & \qquad+ \gamma_2(L, 3n_o, \bar{H}_\epsilon \bar g)+\gamma_3(L, 3n_o) L \bar{\epsilon}{.}
\end{align}
{Therefore, when $\gamma_1(L, 3n_o)\sqrt{\rho}^L \in (0, 1)$, $s(L, n_o, \bar H_\epsilon\bar g)$ is well defined and nonnegative.
It follows that for every $\bar{s}\geq s(L, n_o, \bar H_\epsilon\bar g)$, $\| x_k-\bar{x}_k\| \leq \bar{s}$,
which completes the induction. }
\qed

\end{document}